\documentclass[twocolumn,prc,showpacs,preprintnumbers,amsmath,amssymb,floatfix,nofootinbib]{revtex4}

\usepackage{graphicx}
\usepackage{dcolumn}
\usepackage{bm}
\usepackage{ulem} 
\usepackage[usenames]{color}

\newcommand{\im}{\text{Im}}

\newcommand{\ket}[1]{| \, #1 \, \rangle}

\begin{document}

\preprint{}

\title{Origin of resonances in the chiral unitary approach}

\author{Tetsuo~Hyodo}
\email{thyodo@ph.tum.de}
\affiliation{%
Physik-Department, Technische Universit\"at M\"unchen, 
D-85747 Garching, Germany 
}%

\affiliation{%
Yukawa Institute for Theoretical Physics,
Kyoto University, Kyoto 606-8502, Japan
}%
 
\author{Daisuke~Jido}
\affiliation{%
Yukawa Institute for Theoretical Physics,
Kyoto University, Kyoto 606-8502, Japan
}%
 
\author{Atsushi~Hosaka}%
\affiliation{%
Research Center for Nuclear Physics (RCNP),
Ibaraki, Osaka 567-0047, Japan
}%

\date{\today}
\begin{abstract}
    We study the origin of the resonances associated with pole singularities
    of the scattering amplitude in the chiral unitary approach. We propose 
    a ``natural renormalization'' scheme using the low-energy interaction 
    and the general principle of the scattering theory. We develop a method 
    to distinguish dynamically generated resonances from genuine quark 
    states [Castillejo-Dalitz-Dyson (CDD) poles] using the natural 
    renormalization scheme and phenomenological fitting. Analyzing physical 
    meson-baryon scatterings, we find that the $\Lambda(1405)$ resonance is 
    largely dominated by the meson-baryon molecule component. In contrast, 
    the $N(1535)$ resonance requires a sizable CDD pole contribution, while 
    the effect of the meson-baryon dynamics is also important.
\end{abstract}

\pacs{14.20.--c, 11.30.Rd, 24.85.+p}


\keywords{Chiral dynamics, Dynamical resonance, CDD pole}

\maketitle

\section{Introduction}\label{sec:intro}

Chiral symmetry is one of the guiding principles for studying hadron physics
based on the underlying theory of QCD. The chiral perturbation 
theory~\cite{Weinberg:1979kz,Gasser:1985gg,Pich:1995bw,Ecker:1995gg,
Bernard:1995dp} enables us to study low-energy hadron dynamics 
systematically. By construction, however, perturbative calculations cannot 
be applied to the system with bound states and/or resonances. For instance, 
the leading order term of the chiral perturbation theory describes well the 
$\pi N$ scattering lengths~\cite{Weinberg:1966kf,Tomozawa:1966jm}, while it 
cannot reproduce either the $\pi N$ scattering amplitude around the $\Delta$
resonance energy or the $\bar{K}N$ scattering length due to the presence of 
the $\Lambda(1405)$ resonance below the threshold. To describe the latter 
system in the chiral effective theory, the resonances can be either 
introduced as elementary fields in the Lagrangian or generated dynamically 
in hadron scattering. In general, they can also mix. The clarification of 
these dynamics is one of issues that we discuss in this paper.

Recent developments in the study of resonance scattering based on chiral 
dynamics have been made; the implementation of the unitarity condition on 
the scattering amplitude leads to the nonperturbative resummation of the 
$s$-channel diagrams, generating the resonance pole in the amplitude 
dynamically. This chiral unitary approach was successfully applied to the 
scattering of the pseudoscalar meson with octet baryons~\cite{Kaiser:1995eg,
Oset:1998it,Krippa:1998us,Oller:2000fj,Lutz:2001yb,Garcia-Recio:2003ks}, 
with pseudoscalar mesons~\cite{Dobado:1997ps,Oller:1997ti,Oller:1997ng,
Oller:1998hw}, with decuplet baryons~\cite{Kolomeitsev:2003kt,
Sarkar:2004jh}, with vector mesons~\cite{Lutz:2003fm,Roca:2005nm}, and with 
heavy flavored hadrons~\cite{Kolomeitsev:2003ac,Guo:2006fu,
Gamermann:2006nm}, thanks to the dominant contribution from the 
model-independent low-energy interaction~\cite{Hyodo:2006yk,Hyodo:2006kg}. 
These studies reproduce many scattering observables as well as the 
properties of the observed resonances.

Despite the remarkable success of the chiral unitary approach, the origin of
the resonances is not well understood, especially for the baryonic sector.
One simply expects that the resonances found in this approach are quasibound
states of a meson and a baryon generated by their two-body interaction. 
Hereafter we call this by the meson-baryon picture of the resonance. This 
picture may be in contrast to the description of resonances as genuine quark
states. Such a state is generally called the Castillejo-Dalitz-Dyson (CDD) 
pole~\cite{Castillejo:1956ed,PR124.264}, which is not generated in the 
dynamics of the meson-baryon scattering, but has some different 
origins.\footnote{Strictly speaking, a pole singularity of scattering 
amplitude for an elementary particle is different from the pole originally 
introduced in Ref.~\cite{Castillejo:1956ed}, which gives a 
%
pole of the inverse amplitude.
The presence of the original CDD pole was later interpreted as an 
independent particle participating in the scattering; see, e.g., 
Ref.~\cite{PR124.264}. We will nevertheless use the term ``CDD pole'' to 
indicate the pole of the elementary particle for simplicity.} The 
importance of the CDD pole in the chiral unitary approach was first pointed 
out and discussed in Ref.~\cite{Oller:1998zr}.

In most cases, the CDD pole is introduced explicitly as an elementary field 
in the chiral perturbation theory~\cite{Lee:1996ku,Krippa:1998ix} or in the
unitarized framework~\cite{Sato:1996gk,Igi:1998gn,Oller:1998zr,Jido:2002zk}.
There are, however, some cases in which the CDD pole contribution is hidden 
in the model parameters. For instance, in $\pi\pi$ scattering, it has been 
argued that the pole for the $\rho$ meson is attributed to contact terms in
the higher order Lagrangian~\cite{Oller:1998hw}, which is known to behave as
a contracted resonance propagator in the chiral perturbation 
theory~\cite{Ecker:1989te,Bernard:1995dp}. Hence, the nature of the $\rho$ 
meson is considered to be of the CDD pole, presumably originated from the 
quark dynamics. This observation is in accordance with the study of the 
large $N_c$ limit and the $N_c$ scaling, where the pole of the $\rho$ meson 
behaves as a $\bar{q}q$ resonance rather than a two-meson quasibound 
state~\cite{Oller:1998zr,Pelaez:2003dy}. 

Furthermore, it is also possible to have both CDD pole and dynamical state
in one system~\cite{Oller:1998zr,Igi:1998gn}. In this case, the two 
components will be mixed in physical states. An example of the mixed 
situation has been studied in Ref.~\cite{Albaladejo:2008qa}. There they 
studied the coupling property of the introduced field, which turned out to 
be similar to the corresponding physical resonance in full amplitude. In 
general, such a comparison of the couplings is useful in studying the origin
of the resonance.

In this paper, we study the origin of the resonances in chiral dynamics, 
paying attention to the renormalization procedure. In the chiral unitary 
approach, we need to introduce renormalization parameters (subtraction 
constants) in order to tame the divergence in loop integrals, which have 
been used to fit experimental data~\cite{Oset:1998it,Inoue:2001ip,
Borasoy:2005ie,Oller:2006jw}. Here we propose a different strategy: 
determining the subtraction constant first to study the structure of the 
resonances. Namely, we investigate whether the baryonic resonances obtained 
in the chiral unitary approach are purely dynamically generated resonances 
by meson-baryon scatterings or they have some components other than the 
dynamical one. For this purpose, we develop a renormalization scheme based 
on purely a theoretical argument to exclude the CDD pole contribution in the
loop function. We introduce the following two requirements: (1) the 
scattering equation shares a common feature with ordinary quantum mechanical
problems based on the Schr\"odinger equation, and (2) the obtained 
scattering amplitude is consistent with the low-energy interaction at a 
certain kinematic point. With these conditions, we determine the value of 
the subtraction constant uniquely for the single-channel scattering system 
without CDD poles. We call this scheme ``natural renormalization,'' which 
specifies a standard value of the subtraction constant. Having this scheme, 
we will discuss the meaning of the subtraction constant, which is different 
from the standard value, in what follows.

Next we consider the scattering amplitude in comparison with experimental 
data, and propose a method to extract the low-energy structure of the 
amplitude in the natural renormalization scheme. From the viewpoint of 
renormalization, we first note that the change of the subtraction constant 
can be absorbed into the change of the interaction kernel, once the 
experimental input is given. If the resonance is dominated by the 
meson-baryon component, experimental data are well reproduced in the natural
renormalization scheme with the interaction kernel without the CDD pole 
contribution. If the experimental amplitude requires a large contribution 
from the CDD pole, one has to introduce its effect either in the subtraction
constant or in the kernel interaction. In one way, we can reproduce 
experimental data by suitably choosing the subtraction constant, but keeping
the interaction kernel unchanged. We find that this phenomenological 
amplitude can be equivalently expressed by the natural value of the 
subtraction constant and the interaction with explicit contribution of the 
CDD pole. In this way, the origin of the resonances can be studied, making 
use of the natural renormalization scheme and the experimental input.

This paper is organized as follows. In Sec.~\ref{sec:ChU}, we describe the 
formulation of the chiral unitary approach for a single channel scattering, 
based on the $N/D$ method. In Sec.~\ref{sec:natural}, we discuss the 
properties of the loop function theoretically in the meson-baryon picture. 
We derive the natural value for the subtraction constant from the 
consistency with the general principle and low-energy interaction. In 
Sec.~\ref{sec:interpret}, we present an interpretation of phenomenological 
fitting to experimental data. From the viewpoint of the renormalization, we
analyze the deviation of the subtraction constant from the natural value. We
then generalize the framework to the coupled-channel scattering problem in 
Sec.~\ref{sec:coupled} and perform numerical analysis in 
Sec.~\ref{sec:numerical} for the strangeness $S=-1$ and $S=0$ meson-baryon 
scatterings. The obtained results are discussed in connection with related 
works in Sec.~\ref{sec:discussion}, and concluding remarks are given in the 
last section.

\section{Chiral unitary approach}\label{sec:ChU}

\subsection{Unitarity and $N/D$ method}

In this section, we present the framework of the chiral unitary approach for
$s$-wave meson-baryon scattering. We first discuss the scattering problem in
a single channel for simplicity. Generalization to the coupled-channel 
scattering will be given in Sec.~\ref{sec:coupled}. We consider the 
scattering of a pseudoscalar meson with mass $m$ from a target baryon with 
mass $M_T$. The $s$-channel two-body unitarity condition for the amplitude 
$T(\sqrt{s})$ can be expressed as
\begin{equation}
    \im T^{-1}(\sqrt{s}) = \frac{\rho(\sqrt{s})}{2} ,
    \label{eq:unitarity}
\end{equation}
where $\rho(\sqrt{s})=2M_T\bar{q}/(4\pi\sqrt{s})$ is the two-body phase 
space of the scattering system with 
$\bar{q}=\sqrt{[s-(M_T-m)^2][s-(M_T+m)^2]}/(2\sqrt{s})$. This is the 
so-called elastic unitarity. Based on the $N/D$ method~\cite{Oller:2000fj}, 
the general form of the scattering amplitude satisfying 
Eq.~\eqref{eq:unitarity} is given by
\begin{align}
    T(\sqrt{s})
    =& \frac{1}{V^{-1}(\sqrt{s})-G(\sqrt{s})} ,
    \label{eq:TChU}
\end{align}
where $V(\sqrt{s})$ is a real function expressing dynamical contributions 
other than the $s$-channel unitarity and will be identified as the kernel 
interaction. $G(\sqrt{s})$ is obtained by the once subtracted dispersion 
relation with the phase-space function $\rho(\sqrt s)$:
\begin{align}
    G(\sqrt{s})
    =& -\tilde{a}(s_0)
    -\frac{1}{2\pi}\int_{s^+}^{\infty}
    ds^{\prime}
    \left(
    \frac{\rho(s^{\prime})}{s^{\prime}-s-i\epsilon}
    -\frac{\rho(s^{\prime})}{s^{\prime}-s_0}
    \right) ,
    \label{eq:Gspectral}
\end{align}
where $s^+=(M_T+m)^2$ is the value of $s$ at the $s$-channel threshold, 
$\tilde{a}(s_0)$ is the subtraction constant at the subtraction point $s_0$.
One can easily verify that the amplitude given in Eqs.~\eqref{eq:TChU} and 
\eqref{eq:Gspectral} satisfies Eq.~\eqref{eq:unitarity}.

Equivalently, the function $G(\sqrt s)$ can be written as the finite part of
the meson-baryon loop function 
\begin{align}
    i\int\frac{d^{4}q}{(2\pi)^{4}}
    \frac{2M_T}{(P-q)^{2}-M_T^{2}+i\epsilon}
    \frac{1}{q^{2}-m^{2}+i\epsilon}
    \label{eq:loop} ,
\end{align}
which is logarithmically divergent. Utilizing the dimensional 
regularization, we obtain the same structure as Eq.~\eqref{eq:Gspectral} up 
to a constant
\begin{align}
    G(\sqrt{s})=&\frac{2M_T}{(4\pi)^{2}}
    \Bigl\{a(\mu)+\ln\frac{M_T^{2}}{\mu^{2}}
    +\frac{m^{2}-M_T^{2}+s}{2s}\ln\frac{m^{2}}{M_T^{2}}
    \nonumber\\
    &+\frac{\bar{q}}{\sqrt{s}}
    [\ln(s-(M_T^{2}-m^{2})+2\sqrt{s}\bar{q})
    \nonumber\\
    &
    +\ln(s+(M_T^{2}-m^{2})+2\sqrt{s}\bar{q}) 
    \nonumber\\
    &
    -\ln(-s+(M_T^{2}-m^{2})+2\sqrt{s}\bar{q})
    \nonumber\\
    &
    -\ln(-s-(M_T^{2}-m^{2})+2\sqrt{s}\bar{q})
    ]\Bigr\} ,
    \label{eq:Gdim}
\end{align}
where $a(\mu)$ is the subtraction constant determined at the renormalization
scale $\mu$. The equivalence is verified by noting that both 
Eqs.~\eqref{eq:Gspectral} and \eqref{eq:Gdim} have the same imaginary part 
and that the real part satisfies the dispersion relation. For a single 
channel, there is only one degree of freedom for the regularization. Here we
set $\mu=M_T$ from now on and simply denote the subtraction constant
$a\equiv a(M_T)$, which plays the role of the ultraviolet cutoff parameter 
of the loop integral. A different choice of $\mu$ shifts $a$ by a constant 
value without affecting the physics.

\subsection{Kernel interaction}

Let us consider the meaning of the function $V(\sqrt{s})$, which governs the
dynamics of the system. In principle, $V(\sqrt{s})$ can be constructed once
all the singularities on the complex energy plane are known. In practice, it
is not possible, and we determine it with the help of chiral symmetry.

Regarding the $G(\sqrt{s})$ function as the meson-baryon loop function, we 
can interpret $T(\sqrt{s})$ in Eq.~\eqref{eq:TChU} as the solution of the 
Bethe-Salpeter equation with the kernel interaction $V(\sqrt{s})$. In the 
chiral unitary approach, it was shown that the off-shell effects can be 
absorbed into the renormalization of the kernel 
interaction~\cite{Oller:1997ti,Oset:1998it}, leading to the algebraic 
solution given in Eq.~\eqref{eq:TChU}, which includes the resummation of the
$s$-channel bubble diagrams. One way to determine the interaction kernel 
$V(\sqrt s)$ is to match the unitarized amplitude $T(\sqrt s)$ with the 
chiral perturbation theory order by order~\cite{Oller:2000fj}. At leading 
order, where loops are absent, $V(\sqrt{s})$ is given by the $s$-wave 
interaction of the Weinberg-Tomozawa (WT) term~\cite{Weinberg:1966kf,
Tomozawa:1966jm}
\begin{align}
    V(\sqrt{s})
    =&V_{\text{WT}}(\sqrt{s}) \nonumber \\
    =&-\frac{C}{2f^2}
    [\sqrt{s}-M_T]
    \sim -\frac{C}{2f^2}\omega
    \label{eq:WTterm} ,
\end{align}
where $C$ is the group theoretical factor whose general form is given in 
Ref.~\cite{Hyodo:2006kg}, and $f$ and $\omega$ are the decay constant and
the energy of the meson, respectively. Based on the matching with the chiral
perturbation theory, one can introduce higher order terms in $V(\sqrt{s})$ 
systematically~\cite{Lutz:2001yb,Hyodo:2002pk,Hyodo:2003qa,Borasoy:2005ie,
Oller:2005ig,Oller:2006jw}.

Here we note that if the CDD pole contribution exists, it should be included
in $V(\sqrt{s})$ except for the pole at infinity which can be included in 
the subtraction constant. This is the prescription of the $N/D$ 
method~\cite{Oller:1998zr}. The effect of the CDD pole can be introduced by 
explicitly adding a resonance propagator in the interaction $V(\sqrt{s})$, 
in such a way that it does not violate the low-energy 
theorem~\cite{Igi:1998gn}. While the higher order terms of the chiral 
Lagrangian may contain the CDD pole contribution 
implicitly~\cite{Ecker:1989te,Bernard:1995dp}, the leading order WT 
term~\eqref{eq:WTterm} is apparently not affected by the $s$-channel 
resonance structure~\cite{Gasser:1979hf}. 

\subsection{Properties of the loop function}

For later convenience, we now recall the general properties of the loop 
function. The loop function $G(\sqrt{s})$ is monotonically decreasing in the
energy region below the threshold $\sqrt{s}\leq M_T+m$~\cite{Hyodo:2006yk,
Hyodo:2006kg}. One can verify it by differentiating the expression in 
Eq.~\eqref{eq:Gspectral} with respect to $\sqrt{s}$:
\begin{equation}
    \frac{dG}{d\sqrt{s}}
    =-\frac{1}{2\pi}\int_{s^+}^{\infty}
    ds^{\prime}
    \frac{\rho(s^{\prime})\sqrt{s}}{(s^{\prime}-s+i\epsilon)^2} ,
    \label{eq:Gderi}
\end{equation}
which is negative for $(0\leq)\sqrt{s}\leq M_T+m$. 

The physical $s$-channel scattering takes place above the threshold 
$\sqrt{s}\geq M_T+m$, which is on the unitarity (right hand) cut. The energy
region below the threshold $\sqrt{s}\leq M_T+m$ corresponds to the bound 
state region of the $s$-channel scattering. In the present formulation of 
the $N/D$ method, we fully take into account the unitarity cut, while the 
contribution from the unphysical (left hand) cut is included through 
order-by-order matching. This means that the crossed diagram in the $u$ 
channel is treated only perturbatively. Our amplitude in Eq.~\eqref{eq:TChU}
therefore should not be extrapolated to the energy region below the mass of 
the target $\sqrt{s}\leq M_T$, where the contributions from the $u$-channel 
diagrams become important.

As for the renormalization procedure of the loop function in 
Eq.~\eqref{eq:loop}, one can equivalently utilize procedures other than the 
dimensional regularization, such as the three-momentum cutoff scheme. On the
one hand, the cutoff scheme provides an intuitive interpretation of the loop
function in connection with the second-order perturbation of quantum 
mechanics. On the other hand, the dimensional regularization is compatible 
with the analyticity of the amplitude, which is suitable for the $N/D$ 
method based on dispersion theory. We will make use of both renormalization 
schemes for the loop function in the following sections.

\section{Natural renormalization condition}
\label{sec:natural}

In this section, we propose the ``natural renormalization scheme,'' which 
provides a suitable description for meson-baryon scattering without the CDD 
pole contribution. Our strategy is to determine theoretically the 
subtraction constant of the loop function in order to study the structure of
the resonances. This is in contrast to the previous studies in which the 
subtraction constant is fitted to data. To determine the value of the 
subtraction constant theoretically, throughout this section, we assume that 
there is no contribution to the intermediate states in the loop function 
from the CDD pole and the amplitude follows the low-energy structure 
required by chiral symmetry. For illustration, the interaction kernel 
$V(\sqrt{s})$ is chosen to be the WT term $V_{\text{WT}}(\sqrt{s})$ given in
Eq.~\eqref{eq:WTterm}, which does not contain the CDD pole contribution. We 
may also consider higher order terms, such as quark mass terms. In this 
case, however, some of the higher order terms are known to contain resonance
contributions. For the loop integral, we first show that the subtraction 
constant has an upper limit for the consistency with the physical 
interpretation of the loop function, which is inferred by familiar quantum 
mechanical problems. Next we consider the matching of the unitarized 
amplitude with the low-energy interaction, and we derive the allowed region 
of the subtraction constant. Combining these two conditions, we determine 
the natural subtraction constant for the dynamical generation of resonances 
in a way consistent with low-energy chiral dynamics. Note that this natural 
renormalization scheme is not aimed to describe an arbitrary meson-baryon 
scattering, but it assumes the absence of the CDD pole contribution in the 
loop function, as we discuss in detail below.

\subsection{Consistency with physical loop function}
\label{subsec:consistency}

Let us first consider the sign of the loop function~\eqref{eq:Gdim} below 
the threshold $\sqrt{s}\leq M_T+m$ where the imaginary part vanishes. In the
meson-baryon picture, we can assume that there are no states below the 
threshold contributing to the loop function as intermediate states. This 
sets up the model space of solving the scattering equation. In this case, 
the loop function should be negative below the threshold. This is 
essentially the same as what happens in the perturbative calculations of the
energy of the lowest state which couples to higher states in a quantum
mechanical system, where the energy correction becomes always negative. 

This condition is automatically satisfied in the cutoff regularization; if 
we introduce a three-momentum cutoff $q_{\rm max}$, the loop function can be
written as
\begin{align}
    G^{3d}(\sqrt{s}) 
%
    =&\frac{2M_T}{(2\pi)^2}
    \int_{0}^{q_{\text{max}}}dq\frac{q^2}{E}\frac{1}{\omega}
    \nonumber \\
    &\times \frac{E+\omega}{(\sqrt{s}-(E+\omega)+i\epsilon)
    (\sqrt{s}+E+\omega)} ,
    \nonumber 
\end{align}
with
\begin{align}
    &\quad E=\sqrt{M_T^2+q^2}
     , \quad \omega=\sqrt{m^2+q^2} .
    \nonumber
\end{align}
This is always negative for $\sqrt{s}\leq M_T+m\leq E+\omega$, irrespective 
of the cutoff momentum $q_{\text{max}}$.

In the dimensional regularization, however, the real part can become 
positive if one takes a large positive value for the subtraction constant 
$a$ in Eq.~\eqref{eq:Gdim}. This can be avoided by introducing an upper 
limit for the subtraction constant. As we discussed in the previous section,
our amplitude can be in principle extrapolated down to $\sqrt{s}=M_T$. Since
the loop function below the threshold is a decreasing function as seen in 
Eq.~\eqref{eq:Gderi}, to make the loop function negative for the relevant 
energy region $\sqrt{s}\geq M_T$, it is sufficient for $G(\sqrt s)$ to have 
the negative value at $\sqrt{s}=M_T$, that is,
\begin{align}
    G(M_T)
    \leq &
    0
    \nonumber , 
\end{align}
which is equivalent to
\begin{align}
    a\leq 
    a_{\text{max}}^{(1)}
    =&
    -\Bigl\{\frac{m^{2}}{2M_T^2}\ln\frac{m^{2}}{M_T^{2}}
    +\frac{m\sqrt{m^2-4M_T^2}}{2M_T^2} 
    \nonumber \\
    &
    \times [\ln(m^{2}+m\sqrt{m^2-4M_T^2})
    \nonumber \\
    &
    +\ln(2M_T^{2}-m^{2}+m\sqrt{m^2-4M_T^2}) 
    \nonumber \\
    &
    -\ln(-m^{2}+m\sqrt{m^2-4M_T^2})
    \nonumber \\
    &
    -\ln(-2M_T^{2}+m^{2}+m\sqrt{m^2-4M_T^2})
    ]\Bigr\} .
    \label{eq:max1}
\end{align}
If the subtraction constant satisfies this condition, the loop function with
dimensional regularization is consistent with the physical requirement in 
the region of the $s$-channel scattering ($M_T\leq \sqrt{s}$).

\subsection{Matching with the low-energy interaction}
\label{subsec:matching}

Next we require the amplitude $T(\sqrt{s})$ to follow the chiral low-energy 
theorem~\cite{Hyodo:2006sw,Hyodo:2007jk,Hyodo:2007yc}. As a result of the 
spontaneous chiral symmetry breaking, the scattering amplitude $T(\sqrt s)$ 
can be expanded in powers of momenta of the pseudoscalar meson at low 
energy. Since we choose $V_{\text{WT}}(\sqrt{s})$ for the interaction kernel
as the leading order term of the chiral perturbation theory, the consistency
of the low-energy theorem can be achieved by matching the full scattering 
amplitude $T(\sqrt{s})$ with the kernel interaction 
$V_{\text{WT}}(\sqrt{s})$ at a certain scale $\sqrt{s} = \mu_m$:
\begin{align}
    T(\mu_m)
    =& V_{\text{WT}}(\mu_m),
    \label{eq:matching}
\end{align}
which is realized when
\begin{align}
    G(\mu_m)
    =& 0 ,
    \label{eq:gzero}
\end{align}
as easily seen in Eq.~\eqref{eq:TChU}. Since the subtraction constant is a 
real number, Eq.~\eqref{eq:gzero} should be satisfied below the threshold 
$\mu_m \leq M_T+m$, otherwise the loop function has an imaginary part. 
On the other hand, the matching scale should not be far below the 
threshold, since the $u$-channel cut lies in the region $\sqrt{s}\leq 
M_T-m$, and the effect of the crossing dynamics becomes important at lower 
energies. Therefore, here we set the lower limit of the matching scale at 
$\mu_m = M_T$, to satisfy Eq.~\eqref{eq:matching} within the $s$-channel 
scattering region. In summary, we impose the matching scale to lie in the 
region
\begin{equation}
    M_T\leq \mu_m \leq M_T+m ,
    \label{eq:region}
\end{equation}
which corresponds to choosing the subtraction constant as
\begin{align}
    a_{\min}^{(2)} &\leq a \leq a_{\max}^{(2)} ,
    \label{eq:aregion}
\end{align}
with
\begin{align}
    a_{\min}^{(2)} &= a_{\max}^{(1)} , \quad
    a_{\max}^{(2)} = -\frac{m}{M_T+m}\ln\frac{m^2}{M_T^2} . \nonumber
\end{align}
The matching condition of Eq.~\eqref{eq:matching} was discussed for $\pi N$ 
scattering in Ref.~\cite{Meissner:1999vr}. It is reasonable to set the 
matching scale $\mu_m$ in this region when respecting the low-energy 
expansion. We note that for on-shell kinematics, the three-momentum is zero 
($p=0$) at $\sqrt{s}=M_T+m$, while it takes an imaginary value for the 
vanishing energy of the 
%
Nambu-Goldstone
boson ($\omega = 0$) where 
$\sqrt{s}=\sqrt{M_T^2-m^2}\sim M_T$. Since the chiral perturbation theory is
valid for small four-momentum $p_{\mu}=(\omega,p)$, the matching scale 
$\mu_m$ should lie around the region~\eqref{eq:region}. In the chiral limit 
$m\to 0$, the range~\eqref{eq:region} reduces into one value $\mu_m = M_T$, 
where $\omega=|p|=0$.

One may consider the correction to Eq.~\eqref{eq:matching} from the higher 
order interaction terms, such as the explicit symmetry breaking (quark 
masses) terms. Once again, our aim is to determine the property of the loop 
function by excluding the CDD pole from it. Therefore, in this case we can 
match the amplitude to the interaction $V$ including the higher order 
corrections by the same condition $G(\mu_m)=0$. The inclusion of the higher 
order terms in the interaction does not change the values of the subtraction
constant in Eq.~\eqref{eq:aregion}. In this case, however, we should note 
the possibility of having the CDD pole contribution in the interaction 
kernel $V$ from the higher order chiral Lagrangian. 

\subsection{Natural value for the subtraction constant}

Based on the physical meaning of the loop function and matching with the 
chiral amplitude at low energy, we have derived two conditions for the 
subtraction constant, Eqs.~\eqref{eq:max1} and \eqref{eq:aregion}, which 
read $a_{\rm min}^{(2)} \le a \le a_{\rm max}^{(1)}$ with 
$a_{\min}^{(2)} = a_{\max}^{(1)}$. This means that the subtraction constant 
$a_{\text{natural}}$ which satisfies both conditions is uniquely fixed by
\begin{align}
    a_{\text{natural}} =a_{\min}^{(2)} = a_{\max}^{(1)} .
    \label{eq:natural}
\end{align}
In terms of the zero of the loop function, this condition is equivalent to 
requiring
\begin{align}
    G(\mu_m)=& 0 , \quad \mu_m = M_T .
    \label{eq:naturalG}
\end{align}
This subtraction constant is compatible with the absence of the CDD pole in 
the loop function, as we will discuss below. It also guarantees the matching
of the scattering amplitude with the chiral low-energy interaction. We note 
once again that the subtraction constant so obtained does not 
necessarily explain experimental data. We have just specified a standard 
value of the subtraction constant. The relation to the phenomenologically 
determined value will be discussed in the next section.

\begin{figure}[tbp]
    \centering
    \includegraphics[width=8cm]{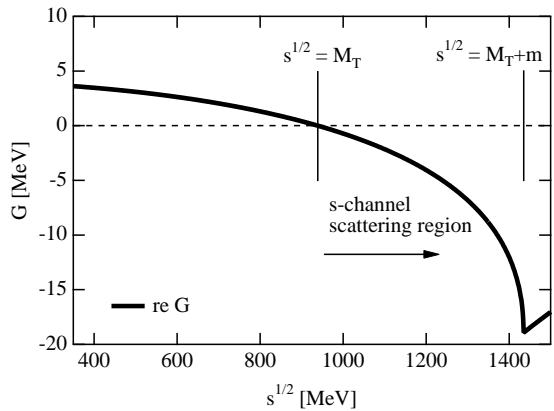}
    \caption{\label{fig:Gex}
    Real part of the loop function $G(\sqrt{s})$ for the $\bar{K}N$ channel 
    when $a=a_{\text{natural}}$ is adopted. The $s$-channel scattering 
    region is defined as $\sqrt{s}>M_T$.}
\end{figure}

In this renormalization condition, we exclude any states below the threshold
as a model space of the unitarization, so that the unitarized amplitude in 
Eq.~\eqref{eq:TChU} naturally implements the meson-baryon picture in the 
model-building. Therefore, with the value of $a_{\text{natural}}$, the loop 
function does not include the CDD pole contribution. The 
condition~\eqref{eq:naturalG} was already proposed in a different context in
Ref.~\cite{Lutz:2001yb}, where the matching with the $u$-channel scattering 
amplitude was emphasized. A similar argument with the present context based 
on chiral symmetry was given in Refs.~\cite{Igi:1998gn,Meissner:1999vr}. Our
point is to regard this condition as the exclusion of the CDD pole in the 
loop function, based on the consistency with the negativeness of the loop 
function. For illustration, the loop function of the $\bar{K}N$ channel with
$a=a_{\text{natural}}$ is plotted in Fig.~\ref{fig:Gex}, where $M_T=939$ 
MeV, $m=496$ MeV, and $f=106.95$ MeV are used.

Let us make some remarks on related works. First, in 
Ref.~\cite{Oller:2000fj}, a ``natural'' value for the subtraction constant 
was estimated to be $a\sim -2$, by comparing the loop function of 
dimensional regularization with that of three-momentum cutoff of $\sim$630 
MeV. This is different from our value of $a_{\text{natural}}$, practically 
and conceptually. In the present context, the value~\eqref{eq:natural} is 
derived for the loop function as unaffected by CDD poles. We used the 
expression of the three-momentum cutoff to illustrate that the real part of 
the loop function is negative below threshold. It is not needed to introduce
the explicit scale (such as $\sim$630 MeV) of the cutoff in our case. The 
``natural'' value in Ref.~\cite{Oller:2000fj} can, in principle, be applied 
to any system, as long as the typical cutoff scale of the physics is around 
$\sim$630 MeV. On the other hand, our natural renormalization scheme is
introduced just for excluding the CDD poles; it does not describe the 
scattering with CDD poles. This possibility is considered in the next 
section. Second, we introduce the condition of matching~\eqref{eq:matching} 
to determine the value of the subtraction constant explicitly, along the 
same line with Ref.~\cite{Meissner:1999vr}. This is different from the 
order-by-order matching introduced in Ref.~\cite{Oller:2000fj}. The latter
is a conceptual matching used to derive the form of the interaction kernel 
$V$. Our condition~\eqref{eq:naturalG} explicitly require the vanishing of 
the loop function at a certain low-energy point. Then the value of the 
subtraction constant is determined, once again, for the loop function 
without CDD poles.

\section{Interpretation of phenomenological model}
\label{sec:interpret}

In this section, we discuss the origin of the dynamically generated 
resonances by reanalyzing the simplest phenomenological model to determine 
the subtraction constant in the chiral unitary approach, in comparison with 
the natural renormalization scheme. Let us assume that we have enough 
experimental data for the system of interest from the low-energy to the 
resonance-energy region. From the viewpoint of the renormalization, once the
scattering amplitude $T$ (observable) is fixed, the change of the 
renormalization parameter in the loop function $G$ can be absorbed into the 
change of the interaction $V$. In other words, we cannot determine {\it a 
priori} the interaction kernel and the loop function separately. Thus, for a
given amplitude $T$, we can construct different sets of interaction $V$ and 
loop function $G$, 
\begin{equation}
   T(\sqrt s) = \frac{1}{V^{-1}(\sqrt s; a) - G(\sqrt s; a)} ,
   \label{eq:ampGene}
\end{equation}
where $a$ labels the renormalization scheme. Once we specify either the loop
function $G$ or the interaction kernel $V$, we also determine the other one 
by Eq.~\eqref{eq:ampGene} to reproduce the same amplitude $T$.

In the conventional phenomenological approaches, the interaction kernel $V$ 
is determined in the beginning by chiral perturbation theory. For instance, 
in the simplest models, the interaction kernel $V$ is chosen to be the 
leading order WT term,
\begin{equation}
   T(\sqrt s) = \frac{1}{V^{-1}_{\text{WT}}(\sqrt s) 
   - G(\sqrt s; a_{\rm pheno})}
   \label{eq:ampPheno} ,
\end{equation}
with the subtraction constant $a_{\text{pheno}}$ in the loop function $G$ 
being a free parameter to reproduce experimental data. We call this 
procedure the phenomenological renormalization scheme. This scheme can 
describe various phenomena well, but the subtraction constant does not 
always satisfy the natural renormalization condition in 
Eq.~\eqref{eq:natural}. Such a subtraction constant takes care of the 
contributions that are not included in the interaction kernel 
$V_{\text{WT}}$. 

The renormalization condition proposed in the previous section is to fix the
subtraction constant such that in the resulting loop function there is no 
contribution from states below the threshold. To achieve the equivalent 
scattering amplitude, we need to adopt a different interaction kernel 
$V_{\rm natural}$ as
\begin{equation}
   T(\sqrt s) = \frac{1}{V_{\rm natural}^{-1}(\sqrt s) - 
   G(\sqrt s; a_{\rm natural})}
   \label{eq:ampNatural} ,
\end{equation}
with the subtraction constant $a_{\rm natural}$ given in 
Eq.~\eqref{eq:natural}. The interaction kernel $V_{\rm natural}$ should be 
modified from $V_{\text{WT}}$ in order to reproduce experimental 
observables.

The physical observable $T$ should equivalently be reproduced by both 
renormalization schemes. Thus, equating the denominators of 
Eqs.~\eqref{eq:ampPheno} and \eqref{eq:ampNatural}
\begin{eqnarray}
    \lefteqn{
    V^{-1}_{\text{natural}}(\sqrt s) - G(\sqrt s; a_{\rm natural}) 
    } \nonumber \\&& 
    =V^{-1}_{\text{WT}}(\sqrt s) - G(\sqrt s; a_{\rm pheno}) \ ,
\end{eqnarray}
we obtain the interaction kernel $V_{\rm natural}$ in the natural 
renormalization scheme as
\begin{equation}
    V^{-1}_{\text{natural}}(\sqrt s) 
    =V^{-1}_{\text{WT}}(\sqrt s) - \frac{2M_{T}}{16\pi^{2}}\Delta a \ ,
\end{equation}
with $\Delta a \equiv a_{\text{pheno}} - a_{\text{natural}} $. Here we have 
exploited the fact that the dependence of $a$ in the loop function $G$ reads
constant shift, as seen in Eq.~\eqref{eq:Gdim}. Using the explicit form of 
the WT term \eqref{eq:WTterm}, we finally obtain the interaction kernel in 
the natural renormalization condition as
\begin{align}
    V_{\text{natural}}(\sqrt s) 
    =&\frac{1}{-\frac{2f^2}{C(\sqrt{s}-M_T)}
    -\frac{2M_T\Delta a}{16\pi^2}}
        \label{eq:Vnatu}  \\
    = &-\frac{8\pi^2}{M_T \Delta a}
    \frac{\sqrt{s}-M_T}{\sqrt{s}-M_{\text{eff}}} ,
    \label{eq:pole} 
\end{align}
with an effective mass
\begin{align}
    M_{\text{eff}}\equiv &
    M_T-\frac{16\pi^2f^2}{CM_T\Delta a} 
    \label{eq:effectivemass} .
\end{align}
Hereafter, we call $V_{\rm natural}(\sqrt{s})$ the effective interaction
in the natural renormalization scheme. The expression in Eq.~\eqref{eq:pole}
tells us that the interaction kernel $V_{\text{natural}}(\sqrt s)$ can have 
a pole, which lies in the $s$-channel scattering region with an attractive 
interaction $C>0$ and a negative value for $\Delta a$. Extracting the WT 
term from the effective interaction~\eqref{eq:pole}, we find 
\begin{align}
    V_{\text{natural}}(\sqrt s) 
    = &-\frac{C}{2f^2}(\sqrt{s}-M_T)+\frac{C}{2f^2}
    \frac{(\sqrt{s}-M_T)^2}{\sqrt{s}-M_{\text{eff}}} 
    \nonumber \\
    \equiv & V_{\text{WT}}(\sqrt s)+\Delta V(\sqrt s;\Delta a) .
    \label{eq:pole2}
\end{align}
The second term can be interpreted as the pole whose energy dependence is 
consistent with the chiral expansion, since the pole term is quadratic in
powers of the meson energy $\omega = \sqrt s - M_{T}$, while the leading WT 
term is linear in it. This is consistent with the schematic discussion made 
in Refs.~\cite{Hyodo:2002pk,Hyodo:2003qa} that the change of the subtraction
constant may introduce the effect of the higher order terms in the kernel 
interaction. Mathematically, it is also possible to have a pole for a 
repulsive interaction $C<0$ with $\Delta a>0$. If the experiments require 
such a value for the phenomenological subtraction constant, the effective 
interaction would be the repulsive contact interaction plus an explicit 
resonance term. The $\pi\pi$ scattering amplitude in the linear $\sigma$ 
model is an example of this case~\cite{Ishida:1996fp}.

The relevance of the second term of Eq.~\eqref{eq:pole2} depends on the 
scale of the effective mass $M_{\rm eff}$, which is obtained by the 
difference of the phenomenological and natural subtraction constants
$\Delta a$. If $\Delta a$ is small, the effective pole mass $M_{\text{eff}}$
becomes large. In this case, the second term of Eq.~\eqref{eq:pole2} can be 
neglected or gives smooth energy dependence in the resonance energy region 
$\sqrt{s} \sim M_{T}+m \ll M_{\text{eff}}$. If the difference $\Delta a$ is
large, the effective mass $M_{\rm eff}$ gets closer to the threshold. In 
this case, the pole contribution is no longer negligible. This means that 
the use of a negative $\Delta a$ with large absolute value is equivalent to 
the introduction of a pole in the chiral Lagrangian. We therefore consider 
that the pole in the effective interaction \eqref{eq:pole} is a source of 
the physical resonances in this case. It was known that the higher order 
term could be a source of a resonance in the full amplitude, because these 
terms behave as the contracted resonance propagator in the $s$ channel. Here
we point out a possible source of the resonance in the conventional chiral 
unitary model, even if we use the leading order chiral interaction.

At this stage, two renormalization schemes~\eqref{eq:ampPheno} and 
\eqref{eq:ampNatural} are interpreted as follows. In the phenomenological 
scheme~\eqref{eq:ampPheno}, the interaction kernel $V_{\text{WT}}$ does not 
include the CDD pole contribution, while in the natural
scheme~\eqref{eq:ampNatural} the loop function $G$ does not contain the CDD 
pole, as discussed in the previous section. Therefore, when the physical 
amplitude contains the CDD pole contribution, the effect is attributed to 
$G(\sqrt s; a_{\rm pheno})$ in the phenomenological scheme, while to 
$V_{\text{natural}}(\sqrt s) $ in the natural scheme. Indeed, we have 
demonstrated that $V_{\text{natural}}(\sqrt s)$ contains a resonance 
propagator. In the limit $\Delta a\to 0$, the two schemes agree with each 
other, which corresponds to the amplitude compatible with the meson-baryon 
picture of resonances, as explained in Sec.~\ref{sec:natural}. Note also 
that in the $N/D$ method, the CDD pole contributions except for those at 
infinity should be included in the interaction kernel $V$, since the loop 
function $G$ expresses the only contribution from the unitarity cut. In this
respect, the phenomenological scheme has accommodated the CDD pole 
contribution in the loop function. In contrast, the natural scheme has more 
similarity to the formulation of the $N/D$ method, as the CDD pole 
contribution is explicitly seen in the interaction kernel.

It is worth noting that the energy dependence of the interaction 
$V_{\text{WT}}(\sqrt{s})$ leads to the pole in the effective interaction,
since the effective pole mass $M_{\text{eff}}$ is obtained by solving the 
equation
\begin{equation}
    1-A\cdot V_{\text{WT}}(\sqrt{s})=0 , \quad
    A=\frac{2M_T\Delta a}{16\pi^2} .
    \nonumber
\end{equation}
Thus, for an energy-independent interaction $V$, no pole can appear. Taking 
into account that the coupling should be a derivative type in the nonlinear 
realization of chiral symmetry, the mechanism can be applied to any 
unitarized model with chiral interaction, such as $\sigma$ and $\rho$ mesons
in the meson-meson scattering.

The interaction kernel in the natural renormalization scheme 
$V_{\text{natural}}(\sqrt s) $ can also be expressed by renormalizing 
$\Delta a$ to an effective coupling strength $f^{\prime}$:
\begin{equation}
    V_{\rm natural}(\sqrt{s})
    \equiv
    -\frac{C}{2(f^{\prime})^2}(\sqrt{s}-M_T) ,
\end{equation}
where the change of the coupling strength is then given by
\begin{align}
    (f^{\prime})^2-f^2
    =&\frac{CM_T\Delta a}{16\pi^2}
    (\sqrt{s}-M_T) .
    \nonumber
\end{align}
In the region of $s$-channel scattering $\sqrt{s}>M_T$ for attractive 
interaction $C>0$, we find that positive $\Delta a$ increases $f^2$ (and the
interaction becomes less attractive), and negative $\Delta a$ decreases 
$f^2$ (more attractive). In this way, we can translate the change of the 
subtraction constant into the change of the strength of the interaction 
kernel. This is again consistent with the argument in 
Refs.~\cite{Hyodo:2002pk,Hyodo:2003qa}. 

As we mentioned, $G(\sqrt{s})$ is monotonically decreasing for $\sqrt{s} 
\leq M_T+m$. Since the subtraction constant $a$ appears with a positive sign
in $G(\sqrt{s})$, we find that positive (negative) $\Delta a$ makes $\mu_m$ 
increase (decrease). In this respect, the renormalization condition 
$\mu_m=M_T$ adopted in Refs.~\cite{Hyodo:2006yk,Hyodo:2006kg} was the most 
advantageous prescription to generating a bound state, with the matching 
scale being in the $s$-channel scattering region $\mu_m \geq 
M_T$~\cite{Hyodo:2006sw,Hyodo:2007jk,Hyodo:2007yc}.

\section{Generalization to the coupled-channel scattering}
\label{sec:coupled}

\begin{figure}[b]
    \centering
    \includegraphics[width=8cm]{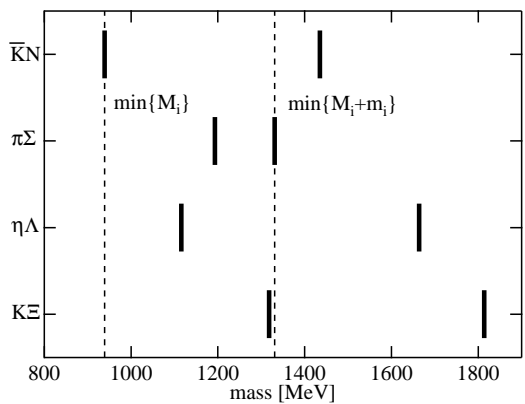}
    \caption{\label{fig:Gcoupled}
    Masses of the target baryons $M_i$ and threshold energies $M_i+m_i$ of 
    channel $i$ for $S=-1$ and $I=0$ meson-baryon scattering. The dashed 
    line on the left denotes the lowest mass of the target hadron 
    ($\text{min}\{M_i\}$), the mass of the nucleon; the dashed line on the 
    right stands for the lowest threshold energy of the $\pi\Sigma$ 
    ($\text{min}\{M_i+m_i\}$).}
\end{figure}

The arguments given so far can be applied to the meson-baryon scattering 
in the flavor-symmetric limit, where channel couplings are absent. In 
practice, the physically interesting system is not flavor symmetric; that 
is, the physical masses for particles break the flavor symmetry. As a 
consequence, we encounter a coupled-channel scattering problem in the chiral
unitary approach. The interaction and amplitude are extended to matrix forms
with channel indices $V_{ij}(\sqrt{s})$ and $T_{ij}(\sqrt{s})$, and the 
scattering equation~\eqref{eq:TChU} is expressed as a matrix equation. The 
loop function is given by a diagonal matrix whose $i$th component is given 
by $G_{i}(\sqrt{s})$, with a different threshold for each channel $i$. The 
generalization of the natural renormalization scheme~\eqref{eq:natural} or 
\eqref{eq:naturalG} to the coupled-channel case is straightforward, once the
differences of the thresholds and the masses of the target hadron $M_i$ are 
properly taken into account. For an illustration of the following argument, 
we show the plot of the mass of the target baryon $M_i$ and the threshold 
$M_i+m_i$ for $S=-1$ and $I=0$ meson-baryon scattering in 
Fig.~\ref{fig:Gcoupled}.

\subsection{Natural values for the subtraction constants \\
in coupled-channel scattering}

We first note that the matching $T_{ij}(\mu_m)=V_{ij}(\mu_m)$ in matrix form
can be achieved when the loop functions in \textit{all} channels are zero at
a common scale $\mu_m$:
\begin{equation}
    G_{i}(\mu_m) =0 .
    \label{eq:Gzerocoupled}
\end{equation}
This equation can be achieved when the imaginary parts of all the loop 
functions vanish, namely, below the lowest threshold:
\begin{align}
     \mu_m\leq \text{min}\{M_i+m_i\}
    \label{eq:upperlimit} .
\end{align}

Recalling the discussion in Sec.~\ref{sec:natural}, 
Eq.~\eqref{eq:Gzerocoupled} should be imposed in the $s$-channel scattering 
region in order to satisfy the consistency with the physical loop function
(Sec.~\ref{subsec:consistency}) and the matching of the full amplitude to 
the low-energy interaction (Sec.~\ref{subsec:matching}). In the 
coupled-channel case, however, the meaning of the ``$s$-channel scattering 
region'' is not clear, since masses of target baryons $M_i$ depend on their 
channel $i$. Here we propose a way to fix the scale $\mu_m$ by
\begin{align}
    G_i(\mu_m)
    = &
    0, \quad \mu_m= \text{min}\{M_i\}
    \label{eq:naturalcouple} ,
\end{align}
in the $s$-channel regions for all the channels. We adopt this condition for
the natural renormalization scheme in the coupled-channel scattering. The 
natural values for the subtraction constants $a_{\text{natural}, i}$ can be 
determined such that the loop function satisfies the 
condition~\eqref{eq:naturalcouple}. With this condition, the loop functions 
in all channels are negative for their $s$-channel scattering region, and 
the full amplitude $T_{ij}(\sqrt{s})$ reduces to the tree level one at 
$\sqrt{s}=\text{min}\{M_i\}$. The scale $\mu_m=\text{min}\{M_i\}$ lies in 
the $u$-channel region for channels with $M_i > \mu_m$, but the 
extrapolation is of the order of mass difference of the particles, which is 
coming from the flavor-symmetry breaking and therefore is not very large.

The condition \eqref{eq:naturalcouple} is one of the ways to achieve the 
natural renormalization in the coupled-channel cases. In principle, we have 
other options for $\mu_m$ that satisfy Eq.~\eqref{eq:Gzerocoupled}. For 
instance, in Ref.~\cite{Lutz:2001yb}, $\mu_m$ is taken to be the mass of the
hadron of the same strangeness as the scattering system, i.e.,
$\mu_m=m_{\Lambda}$ for the $S=-1$ and $I=0$ channels. Here we put more 
weight on the consistency with the physical loop function [$G(\mu_m)\leq 0$]
and choose the lowest mass of the target hadrons with $\mu_m=m_N$.

\subsection{Effective interaction in coupled-channel scattering}

Once the experimental amplitudes are fitted by phenomenological models with 
$a_{\text{pheno}, i}$, we can interpret the origin of the resonances in a
manner similar to that in Sec.~\ref{sec:interpret}. The WT term in the 
coupled-channel case is given by\footnote{Here we ignore the small factor 
$\sqrt{(M_i+E_i)(M_j+E_j)/(4M_iM_j)}$ for simplicity of the discussion of 
the poles. In the numerical analysis in Sec.~\ref{sec:numerical}, we 
include this factor, although the quantitative effect is small: deviation of
pole positions of the scattering amplitude is less than 1 MeV.}
\begin{align}
    V_{\text{WT},ij}(\sqrt{s})
    =&-\frac{C_{ij}}{4f^2}
    [2\sqrt{s}-M_i-M_j]
    \label{eq:WTtermcoupled} ,
\end{align}
with the coupling matrix $C_{ij}$ fixed by the SU(3) group structure of the 
channels. The equation for the amplitude~\eqref{eq:ampGene} should be read 
as a matrix equation. Comparing the phenomenological and natural schemes, 
the effective interaction in the natural renormalization scheme is found to 
be
\begin{equation}
    V_{\text{natural}}(\sqrt{s}) 
    =\left(V_{\text{WT}}^{-1}(\sqrt{s})
    -A\right)^{-1} ,
    \label{eq:effectiveintcouple}
\end{equation}
with a diagonal matrix
\begin{equation}
    A_{ij} = \frac{2M_i\Delta a_i}{16\pi^2}\delta_{ij},
    \quad
    \Delta a_i = a_{\text{pheno}, i}-a_{\text{natural}, i} .
    \nonumber
\end{equation}
Because Eq.~\eqref{eq:effectiveintcouple} is a matrix equation, $\Delta a_i$
in channel $i$ affects the interactions in all channels. To discuss the 
poles in the effective interaction, we rewrite it as 
\begin{align}
    V_{\text{natural}}(\sqrt{s})
    =&V_{\text{WT}}(\sqrt{s})\left(\bm{1}
    -A\cdot V_{\text{WT}}(\sqrt{s})\right)^{-1} 
    \nonumber \\
    =&V_{\text{WT}}(\sqrt{s})\frac{1}
    {\det\left[\bm{1}
    -A\cdot V_{\text{WT}}(\sqrt{s})\right]} \nonumber \\
    &\times
    \text{cof } [\bm{1}-A\cdot V_{\text{WT}}(\sqrt{s})] ,
    \nonumber
\end{align}
where $\det X$ and $\text{cof } X$ are the determinant and the cofactor
matrix of $X$. The poles in the effective interaction are then obtained by
\begin{equation}
    \det\left[\bm{1}
    -A\cdot V_{\text{WT}}(\sqrt{s})\right] =0 .
    \label{eq:polecond}
\end{equation}
As seen in Eq.~\eqref{eq:WTtermcoupled}, each component of 
$V_{\text{WT}}(\sqrt{s})$ is given by the linear function of $\sqrt{s}$, so 
Eq.~\eqref{eq:polecond} is an $n$th order algebraic equation of $\sqrt{s}$ 
for the $n$-channel problem. There are $n$ roots for 
Eq.~\eqref{eq:polecond}, $z_i (i=1,\ldots,n)$ which correspond to poles in 
the effective interaction. It is also possible to have a pair of complex 
poles which are conjugate of each other. We interpret the imaginary part of 
the pole as the width of the pole in the effective interaction, although 
there is no information of the threshold in the construction of the 
effective interaction. For the number of channels smaller than 5, the pole 
positions of the effective interaction can be obtained by analytically 
solving Eq.~\eqref{eq:polecond}.

In the coupled-channel case, around the energy region close to a pole 
position $z_{\text{eff}}$, the effective interaction can be expressed as
\begin{equation}
    V_{\text{natural},ij}(\sqrt{s})
    \sim \frac{g_ig_j}{\sqrt{s}-z_{\text{eff}}} ,
    \nonumber
\end{equation}
where $g_i$ is the coupling strength to channel $i$, which is a complex 
number in general. We can extract $g_i$ from the 
residue of the pole:
\begin{align}
    R_{ij}
    =&(\sqrt{s}-z_{\text{eff}})V_{\text{natural},ij}
    (\sqrt{s})|_{\sqrt{s}=z_{\text{eff}}}
    \nonumber \\
    =& g_ig_j .
    \label{eq:gigj}
\end{align}
When we know all the roots of Eq.~\eqref{eq:polecond}, residues of the pole 
$z_{\text{eff}}$ can be calculated analytically as
\begin{align}
    R_{ij}
    =&V_{\text{WT},im}(z_{\text{eff}})
    \frac{(4f^2)^n}{2^n\text{det}
    [A\cdot C]\Pi_{l\neq \text{eff}}^n(z_{\text{eff}}-z_l)}
    \nonumber 
    \\
    &\times 
    \text{cof }[\bm{1}-A\cdot V_{\text{WT}}(z_{\text{eff}})]_{mj} .
    \nonumber
\end{align}

As in the single-channel case, we define the deviation of the interaction 
$\Delta V_{ij}$ as 
\begin{equation}
    V_{\text{natural},ij}(\sqrt{s})
    =V_{\text{WT},ij}(\sqrt{s})
    +\Delta V_{ij}(\sqrt{s}) ,
    \label{eq:deviation}
\end{equation}
from which we can estimate the effect of $\Delta a_i$ by 
$\Delta V_{ij}(\sqrt{s})$.

\section{Numerical analysis}\label{sec:numerical}

By now we have established the natural renormalization scheme to interpret 
the origin of the poles found in the phenomenological models. In this 
section, we apply our method to physical meson-baryon scatterings in $S=-1$ 
and $I=0$ channel and $S=0$ and $I=1/2$ channel, where the $\Lambda(1405)$ 
and $N(1535)$ resonances are generated, respectively. We use the isospin 
averaged masses for mesons and baryons, and $f=106.95$ MeV. The coupling 
strength $C_{ij}$ can be calculated by the general expression in 
Ref.~\cite{Hyodo:2006kg}, while the explicit numbers can be found in 
Refs.~\cite{Oset:2001cn,Inoue:2001ip}. For these channels, the scattering 
observables such as cross sections and phase shifts are well reproduced by 
the WT term together with the subtraction constants 
$a_{\text{pheno},i}$~\cite{Hyodo:2002pk,Hyodo:2003qa}, which are based on 
the results in Refs.~\cite{Oset:2001cn,Inoue:2001ip}. On the other hand, 
according to Eq.~\eqref{eq:naturalcouple}, we obtain the natural values of 
the subtraction constants $a_{\text{natural},i}$ by setting $G(M_N)=0$ for 
all channels.

Both $a_{\text{pheno},i}$ and $a_{\text{natural},i}$ are shown in 
Table~\ref{tbl:subtractions}. At first glance, the phenomenological 
subtraction constants are similar to the natural values for $S=-1$ channels,
while they are not so for $S=0$ channels. This indicates that 
$\Lambda(1405)$ has a large component of a dynamically generated resonance 
of a meson-baryon system, but $N(1535)$ requires some contribution supplied 
by the subtraction constants, in addition to the dynamical 
meson-baryon component.  

\begin{table}[tbp]
    \centering
    \caption{Natural and phenomenological values~\cite{Hyodo:2003qa}
    for the subtraction constants with the regularization scale 
    $\mu=M_i$.
    \label{tbl:subtractions}}
    \begin{ruledtabular}
    \begin{tabular}{crrrr} 
    $S=-1$ & $\bar{K}N$ & $\pi\Sigma$ 
    & $\eta\Lambda$ & $K\Xi$  \\
    $a_{\text{pheno},i}$ & $-1.042$ & $-0.7228$ 
    & $-1.107$ & $-1.194$  \\
    $a_{\text{natural},i}$ & $-1.150$ & $-0.6995$ 
    & $-1.212$ & $-1.138$  \\
    \hline
    $S=0$ & $\pi N$ & $\eta N$ & $K\Lambda$ & $K\Sigma$  \\
    $a_{\text{pheno},i}$ & 1.509$\phantom{0}$ & $-0.2920$ 
    & 1.454 & $-2.813$  \\
    $a_{\text{natural},i}$ & $-0.3976$ & $-1.239\phantom{0}$
    & $-1.143$ & $-1.138$  \\
    \end{tabular}
    \end{ruledtabular}
\end{table}%

\begin{figure*}[btp]
    \centering
    \includegraphics[width=16cm,clip]{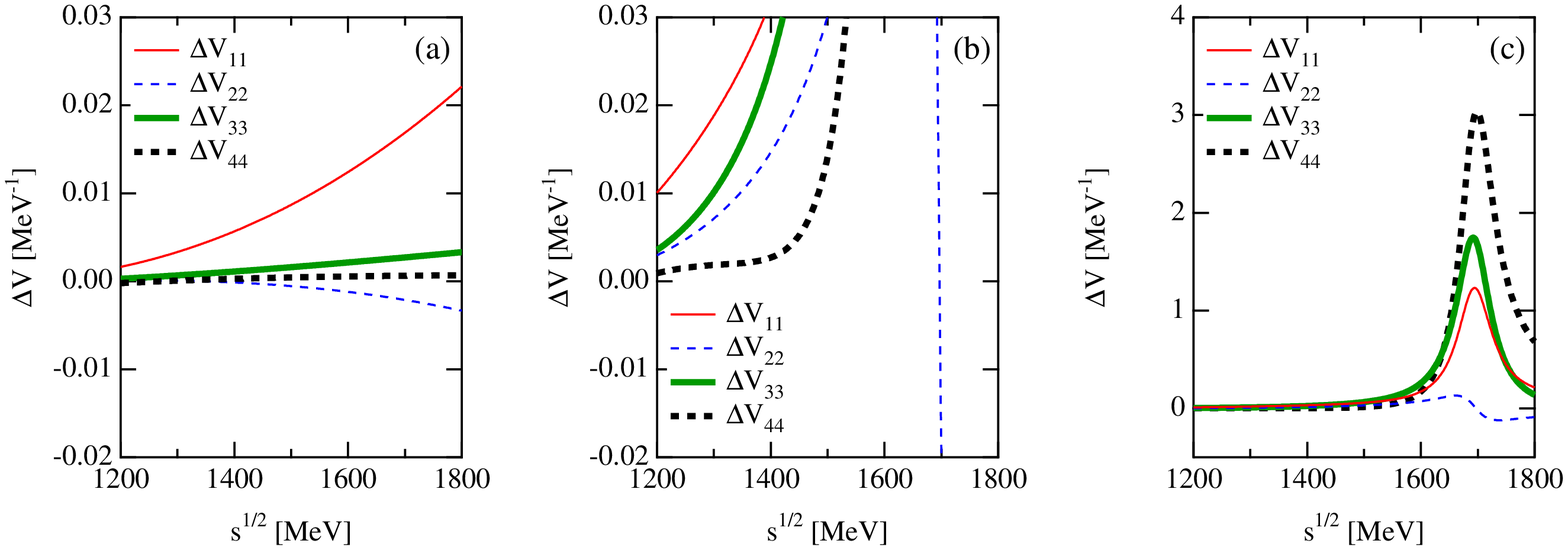}
    \caption{\label{fig:Eff}
    (Color online) Deviations of the effective interactions from the 
    Weinberg-Tomozawa term $\Delta V_{ii}(\sqrt{s})$ defined in 
    Eq.~\eqref{eq:deviation}, (a) $S=-1$ channels, (b) enlargement of panel
    (c), (c) $S=0$ channels. The channels 1--4 correspond to $\bar{K}N$, 
    $\pi\Sigma$, $\eta \Lambda$, and $K\Xi$ for $S=-1$ channels, and to 
    $\pi N$, $\eta N$, $K \Lambda$, and $K\Sigma$ for $S=0$ channels, 
    respectively.}
\end{figure*}%

First of all, we show the pole positions for $\Lambda(1405)$ and $N(1535)$ 
in the amplitudes obtained by the phenomenological renormalization scheme. 
With the phenomenological subtraction constants $a_{\text{pheno},i}$ and the
WT interaction $V_{\text{WT}}$, we find pole positions at
\begin{align}
    z_1^{\Lambda^*} &= 1429 - 14 i \text{ MeV}, \quad
    z_2^{\Lambda^*} = 1397 - 73 i \text{ MeV} ,
    \label{eq:Sm1phen}
\end{align}
for the $\Lambda(1405)$ in $S=-1$ scattering. Note that this resonance is 
expressed by two poles~\cite{Jido:2003cb}, which stem from the attractive 
forces in $\bar{K}N$ and $\pi\Sigma$ channels~\cite{Hyodo:2007jq}. In the 
$S=0$ scattering amplitude, a pole is found at
\begin{align}
    z^{N^*} &= 1493 - 31 i \text{ MeV} ,
    \label{eq:S0phen}
\end{align}
which corresponds to $N(1535)$. These poles reproduce the properties of 
$\Lambda(1405)$ and $N(1535)$ as well as the scattering observables such as 
the total cross sections and the phase shifts~\cite{Hyodo:2002pk,
Hyodo:2003qa}.

\begin{table}[tbp]
    \centering
    \caption{Coupling strengths $g_i$ of the pole in the effective 
    interaction of the $S=0$ channel [Eq.~\eqref{eq:poleeff}].
    \label{tbl:coupling}}
    \begin{ruledtabular}
    \begin{tabular}{ccccc} 
     & $\pi N$ & $\eta N$ & $K\Lambda$ & $K\Sigma$  \\
    \hline
    $g_{i}$ & $3.42+3.18i$ & $-0.192-2.14i$ 
    & $-3.92-4.15i$ & $5.99+4.42i$  \\
    $|g_{i}|$ & $4.67$ & $2.15$ 
    & $5.71$ & $7.44$  \\
    \end{tabular} \\
    \end{ruledtabular}
\end{table}%

Next we evaluate the effective interaction in the natural renormalization 
scheme based on Eq.~\eqref{eq:effectiveintcouple}, and extract the deviation
from the WT term as in Eq.~\eqref{eq:deviation}. We plot the diagonal 
components of the deviation $\Delta V_{ii}(\sqrt{s})$ in Fig.~\ref{fig:Eff}.
We observe that $\Delta V_{ii}(\sqrt{s})$ are small in the $S=-1$ channel 
case, whereas the deviations are large in the $S=0$ channel in the relevant 
energy region of 1400$\leq \sqrt{s}\leq$1600 MeV. Moreover, we observe a 
bump structure at around 1700 MeV in the $S=0$ channel 
[Fig.~\ref{fig:Eff}(c)]. The origin of this structure is due to the poles
found in the effective interaction at
\begin{align}
    z_{\text{eff}}^{N^*} &= 1693 \pm 37 i \text{ MeV}  .
    \label{eq:poleeff} 
\end{align}
These poles may contribute to the structure of the generated $N(1535)$ in 
the full amplitude whose pole is found at a similar energy as shown in 
Eq.~\eqref{eq:S0phen}. We also calculate the coupling strength $g_{i}$ to 
each channel, which is obtained as the residue of the pole in the effective 
interaction as in Eq.~\eqref{eq:gigj}. The values of the couplings are 
summarized in Table~\ref{tbl:coupling}. We observe that the pole strongly 
couples to the $K\Sigma$ channel. As seen in Table~\ref{tbl:subtractions}, 
the difference of the subtraction constants in the $K\Sigma$ channel has a 
large negative value, $\Delta a_{K\Sigma}=-1.67$. This indicates that the 
important ingredient for $N(1535)$ to be added to the WT interaction is in 
the $K\Sigma$ channel.

We estimate theoretical uncertainty of the pole location in 
Eq.~\eqref{eq:poleeff} for $N(1535)$ within the coupled-channel natural 
renormalization scheme. Although we have chosen Eq.~\eqref{eq:naturalcouple}
as a condition for the natural renormalization, as we mentioned above, we 
may choose another matching scale within the region
\begin{align}
    G_i(\mu_m)
    = &
    0, \quad \text{min}\{M_i\}
    \leq \mu_m
    \leq \text{min}\{M_i+m_i\}
    \label{eq:naturalcouple2} ,
\end{align}
in which, except for the original condition $\mu_m=\text{min}\{M_i\}$, loop 
functions in some channels become positive at $\sqrt{s}>M_i$ with the order 
of the flavor-symmetry breaking. Depending on the choice of the natural 
renormalization condition, the values of $a_{\text{natural}}$ change 
slightly. As a consequence, the pole positions, which are the solutions of 
Eq.~\eqref{eq:polecond}, depend on $a_{\text{natural}}$ through the matrix 
$A_{ij}$. Varying the matching scale between the upper and lower 
values of Eq.~\eqref{eq:naturalcouple2}, we find the pole of the effective 
interaction in the region from $z^{N^*}_{\text{eff}} = 1693 \pm 37i$ to
$z^{N^*}_{\text{eff}} = 1673 \pm 146i$ MeV. The pole in the effective 
interaction can be interpreted as a ``bare state,'' which will be dressed by
the meson-baryon cloud through the unitarization procedure. It is therefore 
expected that the pole in the physical amplitude of Eq.~\eqref{eq:S0phen} 
evolves from one of the bare poles found here.

In general, the effective interaction contains $n$ poles, since 
Eq.~\eqref{eq:polecond} has $n$ roots. The relevant point is the energy 
scale of the pole position. If poles appear in the energy region of our
interest, as in the case of $N(1535)$, the effect of the pole on the 
phenomenology is significant. On the other hand, if poles are located away 
from the physically resonant region, these poles are irrelevant to the 
physical observables. In this respect, it is instructive to evaluate the 
pole of the effective interaction for $\Lambda(1405)$. Calculating 
Eq.~\eqref{eq:polecond} for the $S=-1$ channel, we find a pole with almost 
no imaginary part,
\begin{equation}
    z_{\text{eff}}^{\Lambda^*} \sim 7.9 \text{ GeV} .
    \nonumber
\end{equation}
This is far from the relevant energy scale; therefore, the pole plays 
essentially no role for the $\Lambda(1405)$ physics of our interest. Even if
the poles in the physical amplitude of Eq.~\eqref{eq:Sm1phen} originates in 
this bare pole, a substantial effect from the meson-baryon dynamics would be
required. Therefore $\Lambda(1405)$ is largely dominated by the component of
the dynamical meson and baryon.

\begin{figure}[tbp]
    \centering
    \includegraphics[width=8cm,clip]{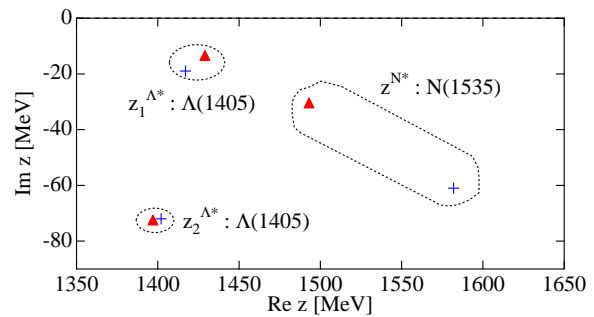}
    \caption{\label{fig:pole}
    (Color online) Pole positions of the meson-baryon scattering amplitudes.
    The triangles stand for the pole positions with the phenomenological 
    amplitude; crosses denote the pole positions in the natural 
    renormalization with the WT interaction. $z_1^{\Lambda^*}$ and 
    $z_2^{\Lambda^*}$ are the poles for $\Lambda(1405)$ in the $S=-1$ 
    scattering amplitude, and $z^{N^*}$ is the pole for $N(1535)$ in the 
    $S=0$ amplitude.}
\end{figure}

We also investigate the pole positions with the natural renormalization with
the WT interaction to see effects of the dynamical component on the
resonance. When we choose the natural values $a_{\text{natural},i}$, we find
\begin{align*}
    z_1^{\Lambda^*} &= 1417 - 19 i \text{ MeV}, \quad
    z_2^{\Lambda^*} = 1402 - 72 i \text{ MeV}  , 
\end{align*}
for $\Lambda(1405)$, and 
\begin{align}
    z^{N^*} &= 1582 - 61 i \text{ MeV}  ,
    \label{eq:S0natural}
\end{align}
for $N(1535)$.\footnote{In Ref.~\cite{Garcia-Recio:2003ks}, they used the WT
term with natural values of the subtraction constants were used, and found a
pole at position similar to that in Eq.~\eqref{eq:S0phen}. We confirm their 
result with $f=90$ MeV; however, the amplitude is not fitted to the 
scattering data. If we adopt the original model in Ref.~\cite{Inoue:2001ip},
namely, by choosing the channel-dependent $f_i$, the pole position of the 
phenomenological model becomes $z^{N^*} = 1533 - 37 i \text{ MeV}$, which is
closer to the result with the natural scheme [Eq.~\eqref{eq:S0natural}], but
not in as good agreement as in the case of the $S=-1$ channel.} We plot the 
pole positions in Fig.~\ref{fig:pole}. The poles for $\Lambda(1405)$ are 
very similar to those obtained by the phenomenological subtraction 
constants. This again indicates the dominance of the meson-baryon component 
in $\Lambda(1405)$. On the other hand, the pole for $N(1535)$ moves to the 
higher energy when we use the natural values. Since a sizable attractive 
interaction exists, a pole can be generated in $S=0$ scattering, although 
the amplitude is not in good agreement with experimental data, as indicated 
by the difference of the pole positions. For the theoretical ambiguity of 
the pole position of Eq.~\eqref{eq:S0natural} in the natural 
renormalization, Eq.~\eqref{eq:naturalcouple2} leads to the pole position of
the amplitude as
\begin{align}
    z^{N^*} &\sim 
    (1582\text{--}1602) \pm (61\text{--}65)i\text{ MeV} 
    .
    \label{eq:znaturalregion}
\end{align}
These results in different natural schemes are still far from the value in 
Eq.~\eqref{eq:S0phen}, with which the amplitude successfully reproduces 
experimental data. 

Study of the coupling properties of the pole for $N(1535)$ is instructive to
further understand the origin of the resonance. In 
Table~\ref{tbl:couplingpheno}, we show the coupling strengths of the 
$N(1535)$ pole in the phenomenological amplitude of Eq.~\eqref{eq:S0phen}.
In Ref.~\cite{Albaladejo:2008qa}, the pole in the physical amplitude 
exhibited a similar coupling tendency with the pole in the effective 
interaction. Based on this observation, Ref.~\cite{Albaladejo:2008qa}
concluded that the CDD pole contribution dominates in the physical state. In
the present case, comparing Table~\ref{tbl:couplingpheno} with that in the 
effective interaction (Table~\ref{tbl:coupling}), we find that this is not 
the case for $N(1535)$ in the present model. On the other hand, the coupling
property of the phenomenological $N(1535)$ is more similar to that of the 
pole in the amplitude by the WT term with the natural renormalization scheme
[Eq.~\eqref{eq:S0natural}] shown in Table~\ref{tbl:couplingnatural}. Since 
the latter is attributed to the meson-baryon dynamical component of the 
resonance, the analysis of the coupling strengths indicates the importance 
of the meson-baryon component in $N(1535)$ in addition to the CDD pole 
contribution.

\begin{table*}[t]
    \centering
    \caption{
    Coupling strengths $g_i$ of the pole in the phenomenological amplitude 
    of $S=0$ channel [Eq.~\eqref{eq:S0phen}].
    \label{tbl:couplingpheno}}
    \begin{ruledtabular}
    \begin{tabular}{ccccc} 
     & $\pi N$ & $\eta N$ & $K\Lambda$ & $K\Sigma$  \\
    \hline
    $g_{i}$ & $0.911+0.256i$ & $1.60-0.374i$ 
    & $-1.40-0.393i$ & $2.92-0.451i$  \\
    $|g_{i}|$ & $0.949$ & $1.64$ 
    & $1.45$ & $2.96$  \\
    \end{tabular} \\
    \end{ruledtabular}
\end{table*}%

\begin{table*}[tb]
    \centering
    \caption{
    Coupling strengths $g_i$ of the pole in the natural renormalization of 
    $S=0$ channel [Eq.~\eqref{eq:S0natural}].
    \label{tbl:couplingnatural}}
    \begin{ruledtabular}
    \begin{tabular}{ccccc} 
     & $\pi N$ & $\eta N$ & $K\Lambda$ & $K\Sigma$  \\
    \hline
    $g_{i}$ & $0.126+0.330i$ & $-1.99-0.700i$ 
    & $-1.63+0.508i$ & $-2.90+0.359i$  \\
    $|g_{i}|$ & $0.353$ & $2.11$ 
    & $1.71$ & $2.93$  \\
    \end{tabular} \\
    \end{ruledtabular}
\end{table*}%

In summary for the numerical analysis, we have studied the origin of 
$\Lambda(1405)$ and $N(1535)$ based on the natural renormalization scheme 
and phenomenological amplitude of the meson-baryon scattering. The $S=-1$ 
scattering and $\Lambda(1405)$ are well reproduced by the natural
renormalization with the WT term, indicating that the $\Lambda(1405)$ 
resonance is a (mostly) pure dynamical resonance. In contrast, the $S=0$ 
scattering and the $N(1535)$ resonance is not reproduced by the WT term 
only, and the translation of the phenomenological subtraction constants into
the low-energy effective interaction requires a pole term of which the mass
is around 1700 MeV in addition to the WT interaction. At the same time, the 
dynamical component is also important for the structure of $N(1535)$, since 
the attractive interaction of the WT term is strong enough to generate a 
resonance in the natural renormalization, and the coupling property of 
$N(1535)$ is closer to the dynamical resonance. Therefore, we interpret 
$N(1535)$ as mixture of a pole singularity of genuine state with the 
dynamical component.

\section{Discussion}\label{sec:discussion}

The results of the present analysis can be argued in various theoretical 
perspectives. There are several discussions about the structure of the 
baryon resonances: three-quark versus five-quark, or hadronic molecule 
versus quark originated structure. In principle, all these structures 
eventually stem from QCD dynamics and mix with each other. Nevertheless, it 
helps our physical understanding to extract several components out of a 
resonance state and inspect the dominant contribution to the resonance. For 
instance, the $\Lambda(1405)$ resonance can be schematically decomposed 
as~\cite{Hyodo:2008ek}
\begin{align}
    \ket{\Lambda(1405)}
    =&N_{MB}\ket{B}\ket{M}
    +\cdots\nonumber \\
    &+N_3\ket{qqq}
    +N_5\ket{qqqq\bar{q}}+\cdots 
    \label{eq:decomp} ,   
\end{align}
where $\ket{B}\ket{M}$ is the dynamical meson-baryon component in the 
scattering theory of hadrons~\cite{PR137.B672,Morgan:1992ge}; and the rest, 
which corresponds to the CDD pole and is not represented by the meson-baryon
state, is expanded by the number of quarks.

According to the decomposition~\eqref{eq:decomp}, the present analysis for 
$\Lambda(1405)$ unveils a large weight of the $N_{MB}$. Probably, the best 
way to disentangle the dynamical component from the CDD pole contribution is
the model-independent determination proposed in Ref.~\cite{Baru:2003qq}. 
Unfortunately, the applicability of this method is limited, and it seems to 
be difficult to deal with the resonances considered in this 
paper~\cite{Hanhart:2007cm}. Therefore, our analysis, though studied in a 
specific model, can be regarded as an alternative approach to this subject 
with larger applicability.

Another powerful method for clarifying the internal structure of the 
resonances is the use of the number of colors $(N_c)$. It is well known that
the only $\bar{q}q$ meson survives in the large $N_c$ limit. The property of
the meson resonances in the large $N_c$ limit was studied in a dynamical 
approach~\cite{Oller:1998zr}. It was found that the $\rho$ meson survives in
the large $N_c$ limit while the $\sigma$ disappears, indicating the 
$\bar{q}q$ nature of the former resonance. A systematic study of the $N_c$ 
scaling of the resonance parameters around $N_c=3$ was performed in 
Ref.~\cite{Pelaez:2003dy}, leading to the same conclusion for the properties
of the mesonic resonances. In the context of the baryon resonance, the 
scaling behavior of the $qqq$ baryon with $N_c$ is known from the general 
argument, so it is possible to investigate whether the $N_3$ component 
dominates. The method of $N_c$ scaling has been applied to $\Lambda(1405)$ 
in Ref.~\cite{Hyodo:2007np}, where the $N_c$ behavior of both poles for 
$\Lambda(1405)$ given in Eq.~\eqref{eq:Sm1phen} indicates their non-$qqq$ 
structure. Concerning $\Lambda(1405)$, the present result ($N_{MB}$ 
dominates) and the result in Ref.~\cite{Hyodo:2007np} ($N_3\ll 1$) 
consistently imply that the $\Lambda(1405)$ resonance is dominated by the 
meson-baryon molecular component.

As for $N(1535)$, we have found substantial contribution other than those 
from $N_{MB}$. There is an interesting possibility of the origin of this CDD
pole contribution: a chiral partner of the ground state nucleon. The chiral 
partner is a parity pair of the particles which transform each other under 
the linear realization of the chiral transformation and become degenerate 
when chiral symmetry is restored. Familiar candidates are $(\rho,a_{1})$ and
$(\sigma, \pi)$ in the meson sector. Since $N(1535)$ is the lowest 
negative-parity state having the same quantum number as the ground state 
nucleon, it is a candidate for the chiral partner of the 
nucleon~\cite{Detar:1988kn,Jido:1998av,Jido:2001nt,Jido:2000nt,Jido:1999hd}.
In the linear realization of chiral symmetry, the chiral partner is 
introduced as an explicit field in the chiral symmetric Lagrangian. Such an
explicit field is expressed as a CDD pole in the chiral unitary approach.
Therefore, the CDD pole found here could be interpreted as the chiral 
partner of the nucleon. 

On the other hand, as indicated by the coupling-strength analysis, the 
strong meson-baryon interaction in the $S=0$ channel also provides a sizable
meson-baryon component on top of the quark-originated $N(1535)$. In 
Ref.~\cite{Jido:2007sm}, electrotransition form factors of $N(1535)$, 
namely, the helicity amplitudes $A_{1/2}$ and $S_{1/2}$, have been discussed
in the meson-baryon picture. There $N(1535)$ is expressed by the chiral 
unitary approach with the phenomenological renormalization scheme and the 
transition $\gamma^* N \to N(1535)$ was computed by considering the photon 
coupling only to the constituent meson and baryon in $N(1535)$. Then 
helicity amplitudes were fairly reproduced, and the ratio 
$A^n_{1/2}/A^p_{1/2}$ agreed well with experimental data. The success of 
this calculation without the photon coupling to the possibly 
quark-originated pole term in the effective interaction implies that the 
meson-baryon components of $N(1535)$ are essential for the structure of 
$N(1535)$ proved by low-energy virtual photon.  

It is instructive to recall the study of exotic hadrons in the chiral 
unitary approach~\cite{Hyodo:2006yk,Hyodo:2006kg} where the natural 
renormalization scheme was adopted. It turned out that the attractive 
interaction of the WT term in exotic channels is not strong enough to 
generate a bound state in the SU(3) limit. As emphasized in the present 
paper, the natural renormalization scheme, together with the WT term as the 
interaction kernel, excludes the CDD pole contribution in the scattering 
amplitude. Thus, the conclusion of Refs.~\cite{Hyodo:2006yk,Hyodo:2006kg} 
is the absence of the $s$-wave exotic hadrons which are dynamically 
generated by a meson and baryon without the CDD pole contribution.

\section{Conclusions}\label{sec:conclude}

We have performed a detailed study of the formulation of the chiral unitary 
approach in order to understand the origin of baryon resonances. We point 
out that a certain choice for the subtraction constants in the dimensional 
regularization leads to the positive value of the loop function below 
threshold. Avoiding this and matching the amplitude with the low-energy 
interaction, we construct the ``natural renormalization'' scheme for the 
loop function in which the CDD pole contribution is excluded. We emphasize 
again that this scheme is not always applied to the physical scattering 
system. But rather our aim is to study the structure of the interaction 
kernel, using the natural renormalization scheme as a starting point.

We then consider the physical meson-baryon scattering with experimental 
data. We compare the natural renormalization scheme with the 
phenomenological scheme in which the subtraction constants are fitted to the
experimental data keeping the interaction kernel unchanged. From the 
viewpoint of the renormalization, we show that the same amplitude can be 
expressed by the natural renormalization scheme with an effective 
interaction kernel which exhibits a propagator of an elementary particle. 
This means the necessity of a seed of the resonance in the kernel 
interaction when the subtraction constant differs from the natural value.
This is another mechanism of the CDD pole contribution even if the kernel 
interaction does not include the contracted resonance propagator in the 
low-energy constant. Although both renormalization schemes achieve the same 
scattering amplitude, the natural scheme is suitable for decomposing the 
singularity of the amplitude along the same line as the $N/D$ method.

We analyze the $S=-1$ and $S=0$ meson-baryon scatterings in which the 
$\Lambda(1405)$ and $N(1535)$ resonances are dynamically generated.
Utilizing the phenomenological fitting, we show that $\Lambda(1405)$ can be
generated in the natural renormalization scheme with the Weinberg-Tomozawa 
term, while $N(1535)$ requires substantial correction in addition to the 
leading order chiral interaction, especially a pole singularity at around 
1700 MeV. These facts indicate that $\Lambda(1405)$ can be regarded almost 
purely as a dynamical state of the meson-baryon scattering, while $N(1535)$ 
may have an appreciable component originated from quark dynamics, together 
with the dynamical component as indicated by the coupling properties.

Our analysis can be applied to any system described by the chiral unitary 
approach. We have also emphasized the importance of the phenomenological 
fitting to the data, otherwise we cannot extract the correct low-energy 
structure which is necessary to interpret the origin of the resonance. 
Hence, precise determination of the meson-hadron scattering data will enable
us to further study the properties of hadron resonances.

\begin{acknowledgments}
T.H. thanks the Japan Society for the Promotion of Science (JSPS) for 
financial support.  This work is supported in part by the Grant for 
Scientific Research (Nos.\ 19853800, 20028004, and 19540297) from 
the Ministry of Education, Culture, Sports, Science and Technology (MEXT) of
Japan. This work was partially done under the Yukawa International Program 
for Quark-Hadron Sciences.
\end{acknowledgments}


%

\end{document}